\documentclass{elsart}
\usepackage{natbib}
\usepackage{epsfig}
\begin{document}
\runauthor{Cicero, Caesar and Vergil}
\begin{frontmatter}

\title{NUCLEAR HALF-LIVES FOR ALPHA RADIOACTIVITY OF ELEMENTS WITH 100 $\leq$ Z $\leq$ 130}

\author[SINP]{P. ROY CHOWDHURY\thanksref{X}} and
\author[SINP,UR,VCU]{C. SAMANTA\thanksref{Y}}

\address[SINP]{Saha Institute of Nuclear Physics, 1/AF Bidhan Nagar, Kolkata 700 064, India }
\address[UR]{Physics Department, Gottwald Science Center, University of Richmond, Richmond, VA 23173, U.S.A.}
\address[VCU]{Physics Department, Virginia Commonwealth University, Richmond, VA 23284-2000, U.S.A.}

and

\author[VECC]{D.N. BASU\thanksref{Z}}
\address[VECC]{Variable  Energy  Cyclotron  Centre, 1/AF Bidhan Nagar, Kolkata 700 064, India }

\thanks[X]{E-mail: partha.roychowdhury@saha.ac.in}
\thanks[Y]{E-mail: chhanda.samanta@saha.ac.in}
\thanks[Z]{E-mail: dnb@veccal.ernet.in}

\begin{abstract}

      Theoretical estimates for the half lives of about 1700 isotopes of heavy elements with 100 $\leq$ Z $\leq$ 130 are tabulated using theoretical $Q$-values. The quantum mechanical tunneling probabilities are calculated within a WKB framework using microscopic nuclear potentials. The microscopic nucleus - nucleus potentials are obtained by folding the densities of interacting nuclei with a density dependent M3Y (DDM3Y) effective nucleon - nucleon interaction. The $\alpha$-decay half lives calculated in this formalism using the experimental $Q$-values were found to be in good agreement over a wide range of experimental data spanning about twenty orders of magnitude. The theoretical $Q$-values used for the present calculations are extracted from three different mass estimates {\it viz.} Myers-Swiatecki [MS], Muntian-Hofmann-Patyk-Sobiczewski [M] and Koura-Tachibana-Uno-Yamada [KUTY]. \\

\end{abstract}

\end{frontmatter}
\pagebreak

\tableofcontents
\listoffigures
\listoftables
\pagebreak

\section{INTRODUCTION}

      In 1939 Bohr and Wheeler \cite{bw39} studied mechanism of nuclear fission on the basis of the liquid drop model (LDM) which treats the nucleus as a drop of charged liquid without any structure. As long as the surface tension in the drop is larger than the repulsive Coulomb force due to the protons, a potential barrier prevents it from splitting and the LDM suggests that the potential barrier approaches zero when the atomic number Z $>$100~\cite{ogac06,oga07}. This puts an upper limit to the stability of nuclei. In mid sixties the importance of nuclear shell effects in stabilising heavy nuclei was realised and possible existence of superheavy elements was predicted~\cite{So66,ms66,Ni69}. The nuclear shell model predicted that the next magic proton number beyond Z~=~82 would be Z~=~114. For neutron, N~=~184 was predicted to be magic with an appreciable shell gap. Due to double magicity the nucleus with  Z~=~114, N~=~184 was predicted to be the center of an island of stability. Element 114 with neutron number N~=~175 has been discovered so far. Measured lifetime~\cite{og04} of this $^{289}114$ nucleus is rather short ($2.7^{+1.4}_{-0.7}$ s) but it is still far away from the desired magic N = 184. Inclusion of higher orders of deformation~\cite{pa89,pa91,be03} suggested that ground state shell correction energy of deformed nuclei around Z~=~108, N~=~162 are comparable to that of the doubly magic $^{298}114$ nucleus which has zero deformation. In fact discovery of high alpha-decay half life of $^{270}$Hs, the doubly magic deformed nucleus,  provides the first evidence of N~=~162 shell stability~\cite{dv06}.

Modern microscopic nuclear theories suggest that the island~of~stability would be  around  Z~=~120, 124, or, 126 and N~=~184 or, N~=~172~\cite{st06,chn05}. These predictions as well as earlier predictions~\cite{Fi72,Ra74,Mo76} that some of the superheavy nuclei might have lifetimes comparable to the age of the earth ($\sim$4.567 $\times$ 10$^9$ years) have encouraged a world-wide effort to search for superheavy neutron-rich long-lived nuclei in laboratories as well as in nature~\cite{ogac06}. In this context it is imperative to find out whether the predicted wide shell gap (or, shell stabilization of super and hyperheavy nuclei without magic gap) at high Z and N values would lead to high stability of the superheavy nuclei against alpha-decay. In this work we present theoretical estimates for the $\alpha$-decay half lives of about 1700 heavy and superheavy elements with Z~=~100~-~130. 

      If the existing theoretical predictions are correct, elements on the islands of stability should not decay via spontaneous fission. Instead they should undergo $\alpha$-decay. As a result, these elements will leave a clear experimental signature: a daughter nucleus that is lighter than its parent by two protons and two neutrons, followed by a grand daughter nucleus that is lighter by four protons and four neutrons, and so on~\cite{og04,oga04}. Heavy elements with Z~=~107~-~112 have been successfully synthesized at GSI, Darmstadt~\cite{ho04,ho00,ac06}. Isotopes of these elements along with Z~=~113~-~116 and 118 have been synthesized at JINR-FLNR, Dubna~\cite{og04,oga04,oga06} and Z~=~110~-~113 have been produced at RIKEN, Japan~\cite{mo04,MO04,mo07}. 

      Periodic table arranges elements according to their outermost electrons which dictate their chemical properties. Due to presence of large number of protons in superheavy nuclei some orbital electrons move with velocities close to the speed of light. The relativistic effects might significantly alter the order of the elements' electronic orbitals, and thus influence their chemical reactivity. For example superheavy element 112, provisionally named ununbium, was predicted to belong to either group 12 of the periodic table (which includes the familiar transition metals zinc, cadmium and mercury) or, act like radon~\cite{tu07}. The heaviest elements which are chemically characterized so far are seaborgium (element 106), bohrium (element 107), hassium (element 108) and recently, the element 112. The first three behave according to their respective positions in groups 6, 7 and 8 of the periodic table, and now it has been confirmed~\cite{ei07} that the element 112 is very volatile and, unlike the inactive radon, reveals a metallic interaction with the gold surface. These adsorption characteristics establish element 112 as a typical element of group 12. There exists still now strong controversies on the real observation of superheavy nuclei with charge higher than 112~\cite{ar00,lo02,gr05} for which heavy element chemistry would be essential. 

      While the experiments have already presented concrete evidence in support of measurable stability of several nuclei with Z $>$ 100, some recent theoretical calculations in the quantum tunneling model have provided realistic estimation of their $\alpha$-decay half lives. In fact, the $\alpha$-decay modes and lifetimes of medium to light heavy nuclei agree well with the predictions of the half life calculations in a WKB framework with DDM3Y interaction~\cite{Ba03,ba04,prc06,prc07,scb07}. Although the generalized liquid drop model [GLDM] ~\cite{rm01} or with its parameters refitted to improve predictability~\cite{dsr07} are less sophisticated, yet those descriptions provide comparable results in certain cases.

      For the heaviest nuclei also calculations in the above framework provide excellent agreement with the measured half lives of 27 new nuclei with atomic numbers between 106 and 118~\cite{ba04,prc06,prc07,scb07}. The purpose of the present paper is to update the half life estimations into the regions of nuclei far from stability and of SHE using three different mass tables~\cite{ms,Mu01, MU03, Mu03,KUTY} as input data which will be useful for furtherance of experimental studies towards the much envisaged island of stability in superheavy nuclei.  

\section{CALCULATIONAL DETAILS}

\vspace{-0.52cm}

      The $\alpha$-decay half lives are calculated in the framework of quantum mechanical tunneling of an $\alpha$-particle from a parent nucleus~\cite{scb07}. The required nuclear interaction potentials are calculated by double folding the density distribution functions of the $\alpha$-particle and the daughter nucleus with density dependent M3Y effective interaction. The microscopic $\alpha$-nucleus potential thus obtained, along with the Coulomb interaction potential and the minimum centrifugal barrier required for the spin-parity conservation~\cite{Ba03}, form the potential barrier. The half lives of nuclear disintegration process via $\alpha$-particle emissions are calculated  using the WKB approximation for barrier penetrability. Spherical charge distributions have been used for calculating the Coulomb interaction potentials. The $Q$-values of $\alpha$-decay are obtained directly from Muntian-Patyk-Hofmann-Sobiczewski estimates and using atomic mass excesses from the theoretical mass predictions of Myers-Swiatecki and Koura-Tachibana-Uno-Yamada, but for $\alpha$-particle measured~\cite{aw03} atomic mass excess [2.42491565~MeV] is used.

\subsection{Calculation of Theoretical $Q$-values}

      From the energetics point of view, spontaneous emission of $\alpha$-particles is allowed if the released energy 

\begin{equation}
 Q_{th} = M - ( M_\alpha + M_d) = \Delta M - (\Delta M_\alpha + \Delta M_d)
\label{seqn1}
\end{equation}
\noindent
is a positive quantity. In this equation, $M$, $M_\alpha$, $M_d$ and $\Delta M$, $\Delta M_\alpha$, $\Delta M_d$ are the atomic masses and the atomic mass excesses of the parent nucleus, the emitted $\alpha$-particle and the residual daughter nucleus, respectively, all expressed in the units of energy. The released energy which is called the decay $Q$-value is obtained theoretically from the theoretical estimates for the atomic mass excesses~\cite{ms,Mu01,MU03,Mu03,KUTY}.

\subsection{Atomic Masses Used as Input Quantities}

      The theoretical $Q$-values $Q_{th}$ used for the present calculations are extracted from three different mass estimates. All the available 314 $Q_{th}$-values for parent nuclei [102 $\le$ Z $\le$ 120, where Z is the atomic number] from I. Muntian, S. Hofmann, Z. Patyk and A. Sobiczewski (M)~\cite{Mu01,MU03,Mu03} are used for half life calculations. Corresponding $Q_{th}$-values calculated using W.D. Myers and W.J. Swiatecki (MS)~\cite{ms} are also used. All possible $\alpha$-decay modes are predicted using theoretical mass estimates of H. Koura, T. Tachibana, M. Uno and M. Yamada (KUTY)~\cite{KUTY} and corresponding half lives are calculated theoretically. In the present work, only for 100 $\le$ Z $\le$ 130, $Q_{th}$-values and half lives are tabulated. 

\subsection{Selection Criteria of Parent Nuclides}

      All the new 27 nuclei with atomic numbers between 106 and 118 are selected for theoretical half life calculations using theoretical $Q$-values and plotted for comparison with the measured half lives. Calculations using experimental $Q$-values are found to provide excellent estimates~\cite{prc06,prc07,scb07} for the $\alpha$-decay half lives. Since the spin parities of SHE and their $\alpha$-decay daughters are not known while those for many of nuclei with atomic number Z $\le$ 102 are known~\cite{Ba03}, we use a lower cut-off of Z for parent nuclei at 100. The upper cut-off Z for parent nuclei is kept at 130 which is limited by the availability of the theoretical mass excesses. This selection criteria provides about 1700 positive $Q_{th}$-values. For all these nuclei half lives are calculated assuming the minimum centrifugal barrier required for the spin-parity conservation to be zero since the spin parities of SHE and their $\alpha$-decay daughters are not known.  

\subsection{Method of Calculation of $\alpha$-Decay Half Lives of Superheavy Nuclides}

      The half life of a parent nucleus decaying via $\alpha$ emission is calculated using the WKB barrier penetration probability. The barrier penetrability with DDM3Y interaction is used to provide estimates of  $\alpha$-decay half lives for $Z~= 100 - 130$  $\alpha$ emitters. This procedure of obtaining nuclear interaction energy for the $\alpha$ - nucleus interaction is quite fundamental in nature since the strengths of the M3Y interaction were fixed \cite{Be77} by fitting its matrix elements in an oscillator basis to those elements of the G-matrix obtained with the Reid-Elliott soft-core NN interaction and the density dependence (DD) was obtained from nuclear matter calculations \cite{Ba04}. Earlier it was shown that the half lives calculated in this framework were more reliable~\cite{Ba03} than by other methods such as the GLDM~\cite{rm01}, the analytic super-asymmetric fission model~\cite{Po91,Po06} or Viola-Seaborg semi-empirical relationship~\cite{Vi} with constants determined by Sobiczewski, Patyk and Cwiok~\cite{VSS89}. It was also shown~\cite{scb07} that the theoretical $Q$ values $Q^M_{th}$ extracted from the mass formula of Muntian et al.~\cite{Mu01,MU03,Mu03} can reasonably reproduce the experimental data on several SHE. 

\subsubsection{$\alpha$-Decay Half Life for Quantum Tunneling with Microscopic Potential}

      The barrier penetrability $P$ in the improved WKB~\cite{ke35} framework for any continuous (rounded) potential barrier  is given by,
\begin{equation}
 P = 1/ [1 + \exp(K)]
\label{seqn2}
\end{equation}
\noindent
where $K$ is the action integral and the decay constant $\lambda=\nu P$ where $\nu$ is calculated from $E_v=\frac{1}{2}h\nu$, the zero point vibration energy. The zero point vibration energies used in the present calculations are $E_v$ = 0.1045$Q$ for even-even, 0.0962$Q$ for odd Z-even N, 0.0907$Q$ for even Z-odd N, 0.0767$Q$ for odd-odd parent nuclei and are the same as that described in ref. \cite{Po86} immediately after eqn.(4) which were obtained from a fit to a selected set of experimental data on $\alpha$ emitters and includes the shell and the pairing effects. The assault frequency $\nu$ comes out to be $5.481\times10^{20}$ s$^{-1}$ and $7.108\times10^{20}$ s$^{-1}$ for the $\alpha$ decays from nuclei $^{217}100$ and  $^{330}130$, respectively, using theoretical $Q$ values calculated from KUTY masses while that for some new superheavy elements using experimental $Q$ values are listed in ref. \cite{prc06}. The half life can thus be obtained from $T_{1/2}=\ln2/\lambda = [(h \ln2) / (2 E_v)] [1 + \exp(K)]$. The action integral $K$ within the WKB approximation is given by

\begin{equation}
 K = (2/\hbar) \int_{R_a}^{R_b} {[2\mu (E(R) - E_v - Q)]}^{1/2} dR
\label{seqn3}
\end{equation}
\noindent
where the total interaction energy $E(R)$ between the $\alpha$ and the residual daughter nucleus is equal to the sum of the nuclear interaction energy $V_N(R)$, Coulomb interaction energy $V_C(R)$ and the centrifugal barrier. Thus

\begin{equation}
 E(R) = V_N(R) + V_C(R) + \hbar^2 c^2 l(l+1) / (2\mu R^2)
\label{seqn4}
\end{equation}   
\noindent
where $c$ is the velocity of light in the vacuum, the reduced mass (in MeV) $\mu = M_{\alpha} M_d/ (M_{\alpha} + M_d)$ and $V_C(R)$ is the Coulomb potential between the $\alpha$ and the residual daughter nucleus. $R_a$ and $R_b$ are the second and third turning points of the WKB action integral determined from the equations 

\begin{equation}
 E(R_a)  = Q + E_v =  E(R_b)
\label{seqn5}
\end{equation}
\noindent
whose solutions provide three turning points. The $\alpha$-particle oscillates between the first and the second turning points and tunnels through the barrier at $R_a$ and $R_b$. The nuclear interaction potential $V_N(R)$ between the daughter nucleus and the emitted particle is obtained in a double folding model~\cite{sa79} as,

\begin{equation}
 V_N(R) = \int \int \rho_1(\vec{r_1}) \rho_2(\vec{r_2}) v[|\vec{r_2} - \vec{r_1} + \vec{R}|] d^3r_1 d^3r_2 
\label{seqn6}
\end{equation}
\noindent
where $\rho_1$ and $\rho_2$ are the density distribution functions for the two composite nuclear fragments and $v[|\vec{r_2} - \vec{r_1} + \vec{R}|]$ is the effective NN interaction. The density distribution function in case of $\alpha$-particle has the Gaussian form $\rho(r) = 0.4229~{\rm exp}( - 0.7024 r^2)$ whose volume integral is equal to $A_\alpha ( = 4 )$, the mass number of $\alpha$-particle. The matter density distribution for the daughter nucleus can be described by the spherically symmetric Fermi function $\rho(r) = \rho_0 / [ 1 + {\rm exp}( (r-c) / a ) ]$ where the equivalent sharp radius $r_\rho$, the half density radius $c$ and the diffuseness for the leptodermous Fermi density distributions are given by $c = r_\rho ( 1 - \pi^2 a^2 / 3 r_\rho^2 )$, $r_\rho = 1.13 A_d^{1/3}$, $a =$ 0.54 fm and the value of the central density $\rho_0$ is fixed by equating the volume integral of the density distribution function to the mass number $A_d$ of the residual daughter nucleus. 

      The distance $s$ between any two nucleons, one belonging to the residual daughter nucleus and other belonging to the emitted $\alpha$, is given by $s = |\vec{r_2} - \vec{r_1} + \vec{R}|$ while the interaction potential between these two nucleons $v(s)$ appearing in eqn.(6) is given by the factorised DDM3Y effective interaction. The general expression for the DDM3Y realistic effective NN interaction used to obtain the double-folded nucleus-nucleus interaction potential is given by,  
\begin{equation}
  v(s,\rho_1,\rho_2,\epsilon) = t^{M3Y}(s,\epsilon)g(\rho_1,\rho_2)
\label{seqn7}
\end{equation}   
\noindent
where the isoscalar $t_{00}^{M3Y}$ of M3Y interaction potential~\cite{sa79} supplemented by zero range potential is given by the following equation: 

\begin{equation}
 t_{00}^{M3Y}(s, \epsilon) = 7999\frac{\exp( - 4s)}{4s} - 2134\frac{\exp( - 2.5s)}{2.5s} - 276 (1 - 0.005\epsilon)\delta(s)
\label{seqn8}
\end{equation} 
\noindent 
where $\epsilon$ is the energy per nucleon and the isovector term does not contribute if anyone (or, both) of the daughter and emitted nuclei involved in the decay process has  N=Z, N and Z being the neutron number and proton number respectively. Therefore in $\alpha$-decay calculations only the isoscalar term contributes. The density dependence term $g(\rho_1, \rho_2)$ can be factorized~\cite{prc06} into a target term times a projectile term as,

\begin{equation}
 g(\rho_1, \rho_2) = C (1 - \beta \rho_1^{2/3}) (1 - \beta \rho_2^{2/3})
\label{seqn9}
\end{equation}   
where C, the overall normalisation constant, is kept equal to unity and the parameter $\beta=1.6$ fm$^2$ ~\cite{prc06} obtained from the nuclear matter calculation can be related to the mean-free path in the nuclear medium. The $\rho_1$ and $\rho_2$ are the density distributions of the $\alpha$-particle and the daughter nucleus respectively.

\subsubsection{ Viola-Seaborg-Sobiczewski systematics for $\alpha$-decay Half Life }

        The $\alpha$ decay half lives estimated by Viola-Seaborg semi-empirical relationship~\cite{Vi} with constants determined by Sobiczewski, Patyk and Cwiok~\cite{VSS89} is given by

\begin{equation}
 log_{10} [T_{1/2}/sec]  = [a Z + b] [Q/MeV]^{-1/2} + cZ + d + h_{log}
\label{seqn10}
\end{equation}
\noindent
where the half-life $T_{1/2}$ is in seconds, the $Q$-value is in MeV,  $Z$ is the atomic number of the parent nucleus. Instead of using original set of constants by Viola and Seaborg, more recent values 
\begin{equation}
 a=+1.66175,~~~~b=-8.5166,~~~~c=-0.20228,~~~~d=-33.9069
\label{seqn11}
\end{equation}
\noindent 
that were determined in an adjustment taking account of new data for new even-even nuclei~\cite{VSS89} are used. The increased deviations in the neighbourhood of magic numbers present with the constants of the original Viola-Seaborg~\cite{Vi} formula get smoothed out by these constants or using the semi-empirical formula based on fission theory~\cite{PP06}. The quantity $h_{log}$ in eqn.(10) accounts for the hindrances associated with odd proton and odd neutron numbers given by Viola and Seaborg~\cite{Vi}, namely

\begin{eqnarray}
 h_{log} = &&~~0~~~~~~for~Z~even-N~even\nonumber\\
            = && 0.772~~for~Z~odd-N~even\nonumber\\
            = && 1.066~~for~Z~even-N~odd\nonumber\\
            = && 1.114~~for~Z~odd-N~odd
\label{seqn12}            
\end{eqnarray}   
\noindent       
The uncertainties in the calculated half lives due to this semi-empirical approach are smaller than the uncertainties due to errors in the calculated energy release. 

\subsection{Comparison with Experimentally Measured $\alpha$-Decay Lifetimes}

      To study the predictive power of the mass formula, $Q$-values are also calculated using the mass formulae [MS] of Myers-Swiatecki [$Q^{MS}_{th}$] ~\cite{ms} and [M] of Muntian-Patyk-Hofmann-Sobiczewski [$Q^{M}_{th}$] ~\cite{Mu01,MU03,Mu03} and [KUTY] of Koura-Tachibana-Uno-Yamada [$Q^{KUTY}_{th}$] ~\cite{KUTY}.
 
       Fig. 1 shows comparison between the experimental $\alpha$-decay half lives~\cite{og04,oga04,oga06,mo07} and the theoretical estimates with $Q_{th}$-values from the mass formulae of Muntian et al. [$Q^{M}_{th}$] ~\cite{Mu01,MU03,Mu03} and Koura et al. [$Q^{KUTY}_{th}$]~\cite{KUTY}. The latter highly overpredicts in most of the cases. In a few cases for odd-odd or odd-even nuclei the mass formula M underestimates the experimental data. This arises from non-zero $l$ transfers~\cite{Ba03,prc07} not considered in the present calculations.

      Table 1 contains $T_{1/2}$ predictions for the $\alpha$-decay half lives of Z~=102~-~120 calculated in the WKB framework with DDM3Y interaction and estimations~\cite{scb07} of Viola-Seaborg semi-empirical relationship~\cite{Vi} with constants determined by Sobiczewski, Patyk and Cwiok~\cite{VSS89} [VSS]. The $Q$-values are extracted from the mass formulae of Myers-Swiatecki [$Q^{MS}_{th}$]~\cite{ms} and Muntian et al. [$Q^{M}_{th}$]~\cite{Mu01,MU03,Mu03}. Calculations with M do not indicate extra stability at Z = 120 and N = 184, or, existence of a SHE above 102 with half life comparable to the age of the earth while the predictions using masses from Koura et al. at high Z, N are sometimes far more than the age of the universe.

      Table 2 contains $\alpha$-decay half lives of nuclei with Z~=~100~-~130 using theoretical $Q$-values [$Q^{KUTY}_{th}$] from Koura et al. [KUTY] mass estimates~\cite{KUTY}. As shown in Fig. 1 half lives with KUTY mostly over predicts the experimental data at large N. Nevertheless, none of the $\alpha$-decay half life calculations present here indicates extra stability of the Z~=~120 and N~=~172 or, 184.

\subsection{Sources of Experimental Data}

      The experimental data for the $\alpha$-decay energies and measured lifetimes of various SHE are taken from  Yu.Ts. Oganessian et al., JINR-FLNR, Dubna ~\cite{og04,oga04,oga06} and Kosuke Morita and co-workers at the RIKEN laboratory in Tokyo, Japan~\cite{mo04,MO04,mo07}. 

\subsection{Summary}

      We present theoretical estimates for the $\alpha$-decay half lives of about 1700 heavy and superheavy elements with Z~=~100~-~130 in WKB framework with DDM3Y interaction, using theoretical $Q$-values. This formalism has been found to be quite reliable when experimental $Q$-values are used~\cite{prc06,prc07,scb07}. The theoretical $Q$-values are taken from the mass formulae of Myers-Swiatecki [MS]~\cite{ms}, Muntian et al. [M]~\cite{Mu01,MU03,Mu03} and Koura et al. [KUTY]~\cite{KUTY}. The Viola-Seaborg-Sobiczewski [VSS] estimates of $\alpha$-decay half lives using the $Q$-values from Muntian et al. [M]~\cite{Mu01,MU03,Mu03} and Myers-Swiatecki [MS]~\cite{ms} mass estimates are also presented for comparison. The updated half life estimations into the regions of very heavy and of superheavy nuclei  will be useful for further experimental studies in region of superheavy nuclei.
\pagebreak

\begin{figure}[h]
\eject\centerline{\epsfig{file=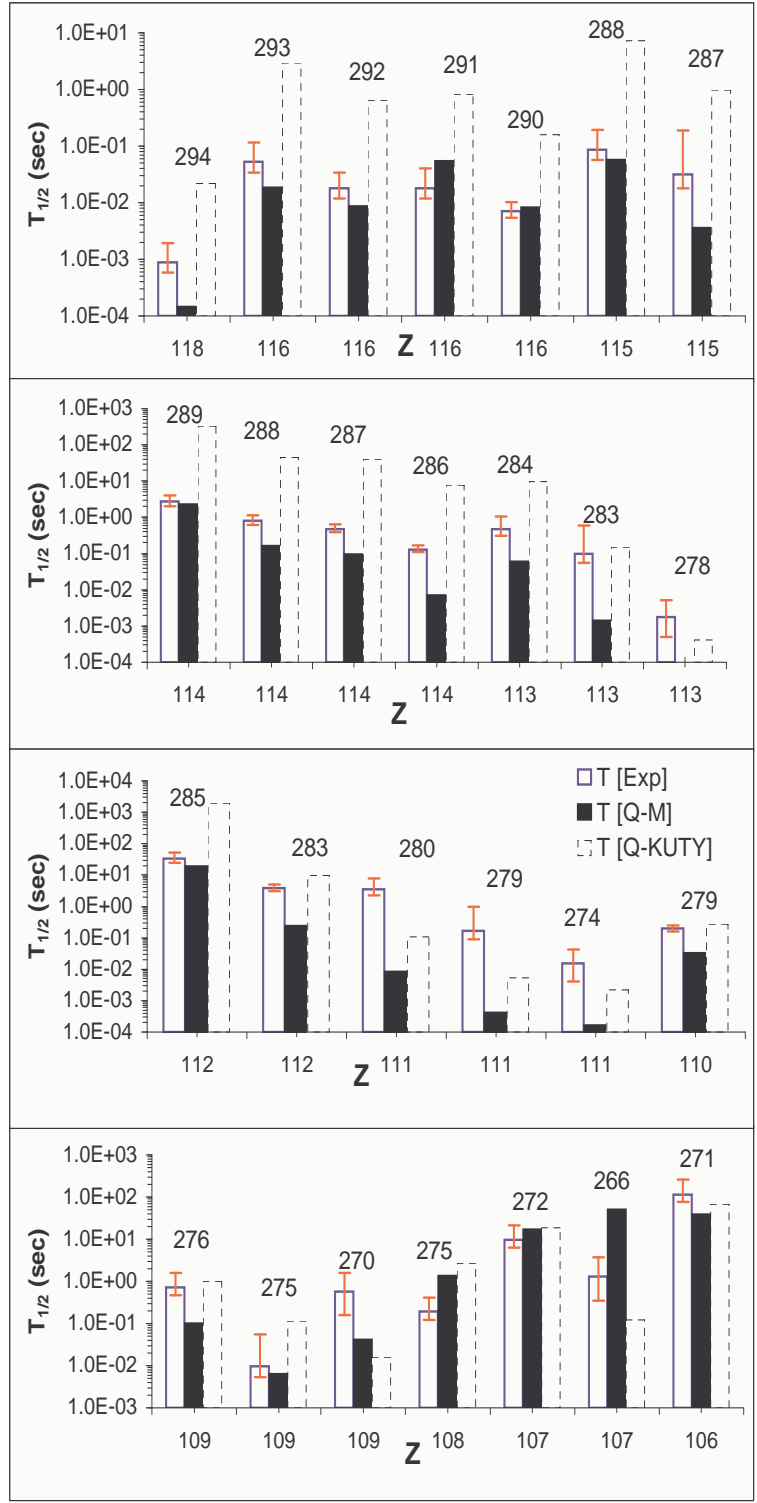,height=20cm,width=10cm}}
\caption
{Plots of $\alpha$-decay half life [T$_{1/2}$(sec)] versus proton number Z for different mass number A (indicated  on top of each coloumn). (a) Hollow columns of solid lines with error bars are experimental $\alpha$-decay half lives (T~[Exp]), (b) filled columns are theoretical half lives ($T~[Q$-$M]$=$T_{1/2}[Q^{M}_{th}]$) in WKB frame work with DDM3Y interaction and [$Q^{M}_{th}$] from Muntian-Patyk-Hofmann-Sobiczewski mass formula, (c) hollow columns of dashed lines are 
($T~[Q$-$KUTY]$=$T_{1/2}[Q^{KUTY}_{th}]$) in the same framework but with [$Q^{KUTY}_{th}$] from Koura-Tachibana-Uno-Yamada mass estimates.}
\label{fig1}
\end{figure}

\pagebreak

\section{EXPLANATION OF TABLES}

      We present theoretical estimates for the $\alpha$-decay half lives of about 1700 heavy and superheavy elements with atomic numbers ranging from 100 to 130 in the WKB framework with DDM3Y interaction, using theoretical $Q$-values in tables 1 and 2. 

\section*{TABLE 1. $\alpha$-decay half lives of nuclei using $Q$-values (MeV) from Muntian-Hofmann-Patyk-Sobiczewski [M] and Myers-Swiatecki [MS]}

      Tabulates for the isotopes of the elements Z=102-120, the $\alpha$-decay $Q$-values calculated using mass esitmates from MS~\cite{ms} and M~\cite{Mu01,MU03,Mu03} and corresponding half lives calculated in WKB framework with DDM3Y interaction and using Viola-Seaborg-Sobeczewiski [VSS] semi-empirical relationship~\cite{VSS89}.
 
\begin{table}[h]
\begin{tabular}{ll} 
Z&     The atomic number of the parent nucleus\\

A&     The mass number of the parent nucleus\\

$Q^{MS}_{th}$&                     Theoretical $Q$-value in MeV obtained using MS mass estimates\\

$Q^M_{th}$&                         Theoretical $Q$-value in MeV obtained using M mass estimates\\

DDM3Y(MS)$\equiv$$T_{1/2}[Q^{MS}_{th}]$&       Half life in seconds calculated using $Q^{MS}_{th}$ and DDM3Y interaction\\ 

DDM3Y(M)$\equiv$$T_{1/2}[Q^M_{th}]$&            Half life in seconds calculated using $Q^M_{th}$ and DDM3Y interaction\\ 

VSS(MS)$\equiv$$T^{VSS}_{1/2}[Q^{MS}_{th}]$&     Half life in seconds estimated by VSS formula using $Q^{MS}_{th}$ \\

VSS(M)$\equiv$$T^{VSS}_{1/2}[Q^M_{th}]$&            Half life in seconds estimated by VSS formula using $Q^M_{th}$ \\

\end{tabular} 
\end{table}

\section*{TABLE 2. $\alpha$-decay half lives of nuclei using $Q$-values (MeV) from and Koura-Tachibana-Uno-Yamada [KUTY]}

      Tabulates for the isotopes of the elements Z=100-130, the $\alpha$-decay $Q$-values calculated using mass esitmates from KUTY~\cite{KUTY} and corresponding half lives calculated in WKB framework with DDM3Y interaction.

\begin{table}[h]
\begin{tabular}{ll} 
Z&     The atomic number of the parent nucleus\\

A&     The mass number of the parent nucleus\\

$Q^{KUTY}_{th}$&                 Theoretical $Q$-value in MeV obtained using KUTY mass estimates\\

$T_{1/2}[Q^{KUTY}_{th}]$&    Half life in seconds calculated using $Q^{KUTY}_{th}$ and DDM3Y interaction\\ 
\end{tabular} 
\end{table}
\pagebreak

\begin{table}[tbh]
\caption{$\alpha$-decay half lives of nuclei using theoretical $Q$-values (MeV) from Muntian-Patyk-Hofmann-Sobiczewski [M] and Myers-Swiatecki [MS] mass estimates.} %\\
%%
 
\end{center}
\end{table}

\begin{table}[tbh]
\caption{$\alpha$-decay half lives of nuclei using theoretical $Q$-values (MeV) from Koura-Tachibana-Uno-Yamada [KUTY] mass estimates.}
\begin{center}
\begin{tabular}{cccccc}
\hline \hline
 Parent &KUTY&DDM3Y(KUTY)&Parent &KUTY&DDM3Y(KUTY)    \\ 
\hline
$^A Z$&$Q^{KUTY}_{th}$&$T_{1/2}[Q^{KUTY}_{th}]$ s&$^A Z$&$Q^{KUTY}_{th}$&$T_{1/2}[Q^{KUTY}_{th}]$ s\\ 
\hline
$^{	217	}	100	$&$	12.495	$&$	2.20E-09	$&	$^{	244	}	100	$&$	8.545	$&$	5.15E-01	$\\								
$^{	218	}	100	$&$	12.535	$&$	8.69E-10	$&	$^{	245	}	100	$&$	8.385	$&$	4.06E+00	$\\								
$^{	219	}	100	$&$	12.405	$&$	2.95E-09	$&	$^{	246	}	100	$&$	8.295	$&$	3.09E+00	$\\								
$^{	220	}	100	$&$	12.355	$&$	1.64E-09	$&	$^{	247	}	100	$&$	8.155	$&$	2.16E+01	$\\								
$^{	221	}	100	$&$	11.375	$&$	2.34E-07	$&	$^{	248	}	100	$&$	8.045	$&$	1.96E+01	$\\								
$^{	222	}	100	$&$	10.985	$&$	6.37E-07	$&	$^{	249	}	100	$&$	7.975	$&$	8.33E+01	$\\
\hline
\hline
\end{tabular} 
\end{center}
\end{table}

\begin{table}[tbh]
%\caption{Alpha-decay half lives of nuclei using $Q$-values (MeV) from KUTY~\cite{KUTY}}
\begin{center}
\begin{tabular}{cccccc}
\hline \hline
 Parent &KUTY&DDM3Y(KUTY)&Parent &KUTY&DDM3Y(KUTY)    \\ 
\hline
$^A Z$&$Q^{KUTY}_{th}$&$T_{1/2}[Q^{KUTY}_{th}]$ s&$^A Z$&$Q^{KUTY}_{th}$&$T_{1/2}[Q^{KUTY}_{th}]$ s\\ 
\hline
$^{	271	}	100	$&$	4.845	$&$	3.31E+17	$&$^{	298	}	100	$&$	0.805	$&$	7.41E+121	$\\
$^{	272	}	100	$&$	4.735	$&$	6.66E+17	$&$^{	299	}	100	$&$	0.585	$&$	8.33E+153	$\\
$^{	273	}	100	$&$	4.635	$&$	1.29E+19	$&$^{	300	}	100	$&$	0.425	$&$	2.15E+188	$\\
$^{	274	}	100	$&$	4.595	$&$	7.89E+18	$&$^{	219	}	101	$&$	12.915	$&$	5.93E-10	$\\
$^{	275	}	100	$&$	4.525	$&$	9.47E+19	$&$^{	220	}	101	$&$	12.905	$&$	1.75E-09	$\\
$^{	276	}	100	$&$	4.365	$&$	6.51E+20	$&$^{	221	}	101	$&$	11.895	$&$	3.49E-08	$\\
\hline
\hline
\end{tabular} 
\end{center}
\end{table}

\begin{table}[tbh]
%\caption{Alpha-decay half lives of nuclei using $Q$-values (MeV) from KUTY~\cite{KUTY}}
\begin{center}
\begin{tabular}{cccccc}
\hline \hline
 Parent &KUTY&DDM3Y(KUTY)&Parent &KUTY&DDM3Y(KUTY)    \\ 
\hline
$^A Z$&$Q^{KUTY}_{th}$&$T_{1/2}[Q^{KUTY}_{th}]$ s&$^A Z$&$Q^{KUTY}_{th}$&$T_{1/2}[Q^{KUTY}_{th}]$ s\\ 
\hline
$^{	243	}	101	$&$	9.175	$&$	2.99E-02	$&$^{	270	}	101	$&$	5.685	$&$	1.02E+13	$\\
$^{	244	}	101	$&$	9.055	$&$	2.30E-01	$&$^{	271	}	101	$&$	5.515	$&$	2.13E+13	$\\
$^{	245	}	101	$&$	8.965	$&$	1.12E-01	$&$^{	272	}	101	$&$	5.155	$&$	2.39E+16	$\\
$^{	246	}	101	$&$	8.845	$&$	9.33E-01	$&$^{	273	}	101	$&$	5.055	$&$	2.14E+16	$\\
$^{	247	}	101	$&$	8.725	$&$	5.62E-01	$&$^{	274	}	101	$&$	4.975	$&$	4.24E+17	$\\
$^{	248	}	101	$&$	8.585	$&$	5.49E+00	$&$^{	275	}	101	$&$	4.905	$&$	2.43E+17	$\\
\hline
\hline
\end{tabular} 
\end{center}
\end{table}

\begin{table}[tbh]
%\caption{Alpha-decay half lives of nuclei using $Q$-values (MeV) from KUTY~\cite{KUTY}}
\begin{center}
\begin{tabular}{cccccc}
\hline \hline
 Parent &KUTY&DDM3Y(KUTY)&Parent &KUTY&DDM3Y(KUTY)    \\ 
\hline
$^A Z$&$Q^{KUTY}_{th}$&$T_{1/2}[Q^{KUTY}_{th}]$ s&$^A Z$&$Q^{KUTY}_{th}$&$T_{1/2}[Q^{KUTY}_{th}]$ s\\ 
\hline
$^{	297	}	101	$&$	1.585	$&$	3.18E+72	$&$^{	244	}	102	$&$	9.595	$&$	2.86E-03	$\\
$^{	298	}	101	$&$	1.445	$&$	7.26E+79	$&$^{	245	}	102	$&$	9.475	$&$	1.38E-02	$\\
$^{	299	}	101	$&$	1.275	$&$	1.90E+87	$&$^{	246	}	102	$&$	9.385	$&$	9.77E-03	$\\
$^{	300	}	101	$&$	1.205	$&$	5.98E+92	$&$^{	247	}	102	$&$	9.275	$&$	4.58E-02	$\\
$^{	301	}	101	$&$	1.045	$&$	1.65E+102	$&$^{	248	}	102	$&$	9.165	$&$	3.75E-02	$\\
$^{	222	}	102	$&$	11.975	$&$	3.03E-08	$&$^{	249	}	102	$&$	9.055	$&$	1.89E-01	$\\
\hline
\hline
\end{tabular} 
\end{center}
\end{table}

\begin{table}[tbh]
%\caption{Alpha-decay half lives of nuclei using $Q$-values (MeV) from KUTY~\cite{KUTY}}
\begin{center}
\begin{tabular}{cccccc}
\hline \hline
 Parent &KUTY&DDM3Y(KUTY)&Parent &KUTY&DDM3Y(KUTY)    \\ 
\hline
$^A Z$&$Q^{KUTY}_{th}$&$T_{1/2}[Q^{KUTY}_{th}]$ s&$^A Z$&$Q^{KUTY}_{th}$&$T_{1/2}[Q^{KUTY}_{th}]$ s\\ 
\hline
$^{	271	}	102	$&$	6.135	$&$	2.87E+10	$&$^{	298	}	102	$&$	1.985	$&$	2.24E+59	$\\
$^{	272	}	102	$&$	5.995	$&$	5.40E+10	$&$^{	299	}	102	$&$	1.925	$&$	8.26E+61	$\\
$^{	273	}	102	$&$	5.625	$&$	2.23E+13	$&$^{	300	}	102	$&$	1.805	$&$	8.70E+64	$\\
$^{	274	}	102	$&$	5.385	$&$	2.21E+14	$&$^{	301	}	102	$&$	1.635	$&$	9.25E+71	$\\
$^{	275	}	102	$&$	5.265	$&$	4.12E+15	$&$^{	302	}	102	$&$	1.505	$&$	2.99E+76	$\\
$^{	276	}	102	$&$	5.275	$&$	1.09E+15	$&$^{	225	}	103	$&$	11.645	$&$	3.88E-07	$\\
\hline
\hline
\end{tabular} 
\end{center}
\end{table}

\include{table1}
\begin{table}[tbh]
%\caption{Alpha-decay half lives of nuclei using $Q$-values (MeV) from KUTY~\cite{KUTY}}
\begin{center}
\begin{tabular}{cccccc}
\hline \hline
 Parent &KUTY&DDM3Y(KUTY)&Parent &KUTY&DDM3Y(KUTY)    \\ 
\hline
$^A Z$&$Q^{KUTY}_{th}$&$T_{1/2}[Q^{KUTY}_{th}]$ s&$^A Z$&$Q^{KUTY}_{th}$&$T_{1/2}[Q^{KUTY}_{th}]$ s\\ 
\hline
$^{	247	}	103	$&$	9.815	$&$	2.61E-03	$&$^{	274	}	103	$&$	6.065	$&$	6.44E+11	$\\
$^{	248	}	103	$&$	9.715	$&$	1.63E-02	$&$^{	275	}	103	$&$	5.695	$&$	1.67E+13	$\\
$^{	249	}	103	$&$	9.595	$&$	9.26E-03	$&$^{	276	}	103	$&$	5.595	$&$	3.39E+14	$\\
$^{	250	}	103	$&$	9.485	$&$	6.65E-02	$&$^{	277	}	103	$&$	5.605	$&$	5.53E+13	$\\
$^{	251	}	103	$&$	9.355	$&$	4.06E-02	$&$^{	278	}	103	$&$	5.415	$&$	4.70E+15	$\\
$^{	252	}	103	$&$	9.235	$&$	3.13E-01	$&$^{	279	}	103	$&$	5.365	$&$	1.83E+15	$\\
\hline
\hline
\end{tabular} 
\end{center}
\end{table}

\begin{table}[tbh]
%\caption{Alpha-decay half lives of nuclei using $Q$-values (MeV) from KUTY~\cite{KUTY}}
\begin{center}
\begin{tabular}{cccccc}
\hline \hline
 Parent &KUTY&DDM3Y(KUTY)&Parent &KUTY&DDM3Y(KUTY)    \\ 
\hline
$^A Z$&$Q^{KUTY}_{th}$&$T_{1/2}[Q^{KUTY}_{th}]$ s&$^A Z$&$Q^{KUTY}_{th}$&$T_{1/2}[Q^{KUTY}_{th}]$ s\\ 
\hline
$^{	301	}	103	$&$	2.235	$&$	1.08E+54	$&$^{	252	}	104	$&$	9.815	$&$	2.97E-03	$\\
$^{	302	}	103	$&$	2.085	$&$	1.09E+59	$&$^{	253	}	104	$&$	9.705	$&$	1.39E-02	$\\
$^{	303	}	103	$&$	1.975	$&$	1.21E+61	$&$^{	254	}	104	$&$	9.585	$&$	1.17E-02	$\\
$^{	228	}	104	$&$	11.395	$&$	1.47E-06	$&$^{	255	}	104	$&$	9.525	$&$	3.99E-02	$\\
$^{	229	}	104	$&$	10.805	$&$	6.19E-05	$&$^{	256	}	104	$&$	9.365	$&$	4.42E-02	$\\
$^{	230	}	104	$&$	10.675	$&$	5.11E-05	$&$^{	257	}	104	$&$	9.315	$&$	1.44E-01	$\\
\hline
\hline
\end{tabular} 
\end{center}
\end{table}

\begin{table}[tbh]
%\caption{Alpha-decay half lives of nuclei using $Q$-values (MeV) from KUTY~\cite{KUTY}}
\begin{center}
\begin{tabular}{cccccc}
\hline \hline
 Parent &KUTY&DDM3Y(KUTY)&Parent &KUTY&DDM3Y(KUTY)    \\ 
\hline
$^A Z$&$Q^{KUTY}_{th}$&$T_{1/2}[Q^{KUTY}_{th}]$ s&$^A Z$&$Q^{KUTY}_{th}$&$T_{1/2}[Q^{KUTY}_{th}]$ s\\ 
\hline
$^{	279	}	104	$&$	5.765	$&$	3.05E+13	$&$^{	232	}	105	$&$	12.695	$&$	4.23E-08	$\\
$^{	280	}	104	$&$	5.695	$&$	2.53E+13	$&$^{	233	}	105	$&$	14.065	$&$	7.67E-11	$\\
$^{	281	}	104	$&$	5.695	$&$	7.57E+13	$&$^{	234	}	105	$&$	13.565	$&$	1.30E-09	$\\
$^{	282	}	104	$&$	5.615	$&$	7.34E+13	$&$^{	235	}	105	$&$	13.185	$&$	1.72E-09	$\\
$^{	283	}	104	$&$	5.555	$&$	5.25E+14	$&$^{	236	}	105	$&$	12.845	$&$	1.95E-08	$\\
$^{	284	}	104	$&$	5.375	$&$	2.33E+15	$&$^{	237	}	105	$&$	12.755	$&$	8.77E-09	$\\
\hline
\hline
\end{tabular} 
\end{center}
\end{table}

\begin{table}[tbh]
%\caption{Alpha-decay half lives of nuclei using $Q$-values (MeV) from KUTY~\cite{KUTY}}
\begin{center}
\begin{tabular}{cccccc}
\hline \hline
 Parent &KUTY&DDM3Y(KUTY)&Parent &KUTY&DDM3Y(KUTY)    \\ 
\hline
$^A Z$&$Q^{KUTY}_{th}$&$T_{1/2}[Q^{KUTY}_{th}]$ s&$^A Z$&$Q^{KUTY}_{th}$&$T_{1/2}[Q^{KUTY}_{th}]$ s\\ 
\hline
$^{	259	}	105	$&$	9.595	$&$	3.64E-02	$&$^{	286	}	105	$&$	5.425	$&$	3.82E+16	$\\
$^{	260	}	105	$&$	9.555	$&$	1.66E-01	$&$^{	287	}	105	$&$	5.265	$&$	8.18E+16	$\\
$^{	261	}	105	$&$	9.405	$&$	1.15E-01	$&$^{	288	}	105	$&$	5.235	$&$	7.59E+17	$\\
$^{	262	}	105	$&$	9.365	$&$	5.58E-01	$&$^{	289	}	105	$&$	5.155	$&$	4.82E+17	$\\
$^{	263	}	105	$&$	9.215	$&$	3.93E-01	$&$^{	290	}	105	$&$	5.935	$&$	2.32E+13	$\\
$^{	264	}	105	$&$	9.005	$&$	6.07E+00	$&$^{	291	}	105	$&$	6.565	$&$	1.90E+09	$\\
\hline
\hline
\end{tabular} 
\end{center}
\end{table}

\begin{table}[tbh]
%\caption{Alpha-decay half lives of nuclei using $Q$-values (MeV) from KUTY~\cite{KUTY}}
\begin{center}
\begin{tabular}{cccccc}
\hline \hline
 Parent &KUTY&DDM3Y(KUTY)&Parent &KUTY&DDM3Y(KUTY)    \\ 
\hline
$^A Z$&$Q^{KUTY}_{th}$&$T_{1/2}[Q^{KUTY}_{th}]$ s&$^A Z$&$Q^{KUTY}_{th}$&$T_{1/2}[Q^{KUTY}_{th}]$ s\\ 
\hline
$^{	241	}	106	$&$	11.905	$&$	8.31E-07	$&$^{	268	}	106	$&$	9.065	$&$	1.27E+00	$\\
$^{	242	}	106	$&$	11.815	$&$	5.41E-07	$&$^{	269	}	106	$&$	8.875	$&$	1.19E+01	$\\
$^{	243	}	106	$&$	11.735	$&$	1.69E-06	$&$^{	270	}	106	$&$	8.725	$&$	1.35E+01	$\\
$^{	244	}	106	$&$	11.685	$&$	9.16E-07	$&$^{	271	}	106	$&$	8.635	$&$	6.79E+01	$\\
$^{	245	}	106	$&$	11.595	$&$	3.04E-06	$&$^{	272	}	106	$&$	8.635	$&$	2.57E+01	$\\
$^{	246	}	106	$&$	11.505	$&$	2.00E-06	$&$^{	273	}	106	$&$	8.385	$&$	4.35E+02	$\\
\hline
\hline
\end{tabular} 
\end{center}
\end{table}

\begin{table}[tbh]
%\caption{Alpha-decay half lives of nuclei using $Q$-values (MeV) from KUTY~\cite{KUTY}}
\begin{center}
\begin{tabular}{cccccc}
\hline \hline
 Parent &KUTY&DDM3Y(KUTY)&Parent &KUTY&DDM3Y(KUTY)    \\ 
\hline
$^A Z$&$Q^{KUTY}_{th}$&$T_{1/2}[Q^{KUTY}_{th}]$ s&$^A Z$&$Q^{KUTY}_{th}$&$T_{1/2}[Q^{KUTY}_{th}]$ s\\ 
\hline
$^{	295	}	106	$&$	5.085	$&$	7.57E+18	$&$^{	252	}	107	$&$	11.425	$&$	2.83E-05	$\\
$^{	296	}	106	$&$	4.895	$&$	5.89E+19	$&$^{	253	}	107	$&$	11.345	$&$	1.20E-05	$\\
$^{	297	}	106	$&$	4.705	$&$	6.58E+21	$&$^{	254	}	107	$&$	11.255	$&$	6.19E-05	$\\
$^{	298	}	106	$&$	4.585	$&$	1.97E+22	$&$^{	255	}	107	$&$	11.165	$&$	2.78E-05	$\\
$^{	299	}	106	$&$	4.425	$&$	1.78E+24	$&$^{	256	}	107	$&$	11.055	$&$	1.61E-04	$\\
$^{	300	}	106	$&$	4.295	$&$	7.54E+24	$&$^{	257	}	107	$&$	10.955	$&$	7.63E-05	$\\
\hline
\hline
\end{tabular} 
\end{center}
\end{table}

\begin{table}[tbh]
%\caption{Alpha-decay half lives of nuclei using $Q$-values (MeV) from KUTY~\cite{KUTY}}
\begin{center}
\begin{tabular}{cccccc}
\hline \hline
 Parent &KUTY&DDM3Y(KUTY)&Parent &KUTY&DDM3Y(KUTY)    \\ 
\hline
$^A Z$&$Q^{KUTY}_{th}$&$T_{1/2}[Q^{KUTY}_{th}]$ s&$^A Z$&$Q^{KUTY}_{th}$&$T_{1/2}[Q^{KUTY}_{th}]$ s\\ 
\hline
$^{	279	}	107	$&$	7.575	$&$	6.64E+05	$&$^{	306	}	107	$&$	4.135	$&$	1.64E+28	$\\
$^{	280	}	107	$&$	7.215	$&$	9.79E+07	$&$^{	307	}	107	$&$	3.995	$&$	6.53E+28	$\\
$^{	281	}	107	$&$	6.925	$&$	4.18E+08	$&$^{	239	}	108	$&$	13.815	$&$	1.38E-09	$\\
$^{	282	}	107	$&$	6.735	$&$	1.59E+10	$&$^{	240	}	108	$&$	13.475	$&$	2.19E-09	$\\
$^{	283	}	107	$&$	6.705	$&$	4.54E+09	$&$^{	241	}	108	$&$	13.175	$&$	1.49E-08	$\\
$^{	284	}	107	$&$	6.685	$&$	2.64E+10	$&$^{	242	}	108	$&$	13.055	$&$	1.06E-08	$\\
\hline
\hline
\end{tabular} 
\end{center}
\end{table}

\begin{table}[tbh]
%\caption{Alpha-decay half lives of nuclei using $Q$-values (MeV) from KUTY~\cite{KUTY}}
\begin{center}
\begin{tabular}{cccccc}
\hline \hline
 Parent &KUTY&DDM3Y(KUTY)&Parent &KUTY&DDM3Y(KUTY)    \\ 
\hline
$^A Z$&$Q^{KUTY}_{th}$&$T_{1/2}[Q^{KUTY}_{th}]$ s&$^A Z$&$Q^{KUTY}_{th}$&$T_{1/2}[Q^{KUTY}_{th}]$ s\\ 
\hline
$^{	264	}	108	$&$	10.775	$&$	2.10E-04	$&$^{	291	}	108	$&$	6.275	$&$	2.71E+12	$\\
$^{	265	}	108	$&$	10.645	$&$	9.79E-04	$&$^{	292	}	108	$&$	6.175	$&$	3.04E+12	$\\
$^{	266	}	108	$&$	10.395	$&$	1.61E-03	$&$^{	293	}	108	$&$	6.945	$&$	1.03E+09	$\\
$^{	267	}	108	$&$	10.325	$&$	5.61E-03	$&$^{	294	}	108	$&$	7.575	$&$	6.21E+05	$\\
$^{	268	}	108	$&$	10.135	$&$	6.85E-03	$&$^{	295	}	108	$&$	7.075	$&$	2.45E+08	$\\
$^{	269	}	108	$&$	10.025	$&$	3.24E-02	$&$^{	296	}	108	$&$	6.735	$&$	3.37E+09	$\\
\hline
\hline
\end{tabular} 
\end{center}
\end{table}

\begin{table}[tbh]
%\caption{Alpha-decay half lives of nuclei using $Q$-values (MeV) from KUTY~\cite{KUTY}}
\begin{center}
\begin{tabular}{cccccc}
\hline \hline
 Parent &KUTY&DDM3Y(KUTY)&Parent &KUTY&DDM3Y(KUTY)    \\ 
\hline
$^A Z$&$Q^{KUTY}_{th}$&$T_{1/2}[Q^{KUTY}_{th}]$ s&$^A Z$&$Q^{KUTY}_{th}$&$T_{1/2}[Q^{KUTY}_{th}]$ s\\ 
\hline
$^{	251	}	109	$&$	12.435	$&$	3.33E-07	$&$^{	278	}	109	$&$	9.525	$&$	3.71E+00	$\\
$^{	252	}	109	$&$	12.345	$&$	1.62E-06	$&$^{	279	}	109	$&$	9.055	$&$	2.38E+01	$\\
$^{	253	}	109	$&$	12.265	$&$	6.58E-07	$&$^{	280	}	109	$&$	8.695	$&$	1.41E+03	$\\
$^{	254	}	109	$&$	12.185	$&$	3.11E-06	$&$^{	281	}	109	$&$	8.295	$&$	7.96E+03	$\\
$^{	255	}	109	$&$	12.125	$&$	1.16E-06	$&$^{	282	}	109	$&$	7.955	$&$	6.26E+05	$\\
$^{	256	}	109	$&$	12.045	$&$	5.54E-06	$&$^{	283	}	109	$&$	7.685	$&$	1.60E+06	$\\
\hline
\hline
\end{tabular} 
\end{center}
\end{table}

\begin{table}[tbh]
%\caption{Alpha-decay half lives of nuclei using $Q$-values (MeV) from KUTY~\cite{KUTY}}
\begin{center}
\begin{tabular}{cccccc}
\hline \hline
 Parent &KUTY&DDM3Y(KUTY)&Parent &KUTY&DDM3Y(KUTY)    \\ 
\hline
$^A Z$&$Q^{KUTY}_{th}$&$T_{1/2}[Q^{KUTY}_{th}]$ s&$^A Z$&$Q^{KUTY}_{th}$&$T_{1/2}[Q^{KUTY}_{th}]$ s\\ 
\hline
$^{	305	}	109	$&$	5.555	$&$	7.99E+16	$&$^{	267	}	110	$&$	11.475	$&$	5.32E-05	$\\
$^{	306	}	109	$&$	5.405	$&$	4.88E+18	$&$^{	268	}	110	$&$	11.375	$&$	3.65E-05	$\\
$^{	307	}	109	$&$	5.305	$&$	4.10E+18	$&$^{	269	}	110	$&$	11.255	$&$	1.53E-04	$\\
$^{	308	}	109	$&$	5.165	$&$	2.46E+20	$&$^{	270	}	110	$&$	11.075	$&$	1.60E-04	$\\
$^{	309	}	109	$&$	5.045	$&$	3.17E+20	$&$^{	271	}	110	$&$	10.865	$&$	1.13E-03	$\\
$^{	245	}	110	$&$	13.585	$&$	9.76E-09	$&$^{	272	}	110	$&$	10.755	$&$	8.36E-04	$\\
\hline
\hline
\end{tabular} 
\end{center}
\end{table}

\begin{table}[tbh]
%\caption{Alpha-decay half lives of nuclei using $Q$-values (MeV) from KUTY~\cite{KUTY}}
\begin{center}
\begin{tabular}{cccccc}
\hline \hline
 Parent &KUTY&DDM3Y(KUTY)&Parent &KUTY&DDM3Y(KUTY)    \\ 
\hline
$^A Z$&$Q^{KUTY}_{th}$&$T_{1/2}[Q^{KUTY}_{th}]$ s&$^A Z$&$Q^{KUTY}_{th}$&$T_{1/2}[Q^{KUTY}_{th}]$ s\\ 
\hline
$^{	294	}	110	$&$	6.835	$&$	9.82E+09	$&$^{	258	}	111	$&$	12.815	$&$	6.87E-07	$\\
$^{	295	}	110	$&$	7.585	$&$	1.15E+07	$&$^{	259	}	111	$&$	12.765	$&$	2.44E-07	$\\
$^{	296	}	110	$&$	8.215	$&$	1.39E+04	$&$^{	260	}	111	$&$	12.665	$&$	1.23E-06	$\\
$^{	297	}	110	$&$	7.875	$&$	7.40E+05	$&$^{	261	}	111	$&$	12.565	$&$	5.58E-07	$\\
$^{	298	}	110	$&$	7.625	$&$	2.57E+06	$&$^{	262	}	111	$&$	12.475	$&$	2.67E-06	$\\
$^{	299	}	110	$&$	7.445	$&$	4.17E+07	$&$^{	263	}	111	$&$	12.345	$&$	1.39E-06	$\\
\hline
\hline
\end{tabular} 
\end{center}
\end{table}

\begin{table}[tbh]
%\caption{Alpha-decay half lives of nuclei using $Q$-values (MeV) from KUTY~\cite{KUTY}}
\begin{center}
\begin{tabular}{cccccc}
\hline \hline
 Parent &KUTY&DDM3Y(KUTY)&Parent &KUTY&DDM3Y(KUTY)    \\ 
\hline
$^A Z$&$Q^{KUTY}_{th}$&$T_{1/2}[Q^{KUTY}_{th}]$ s&$^A Z$&$Q^{KUTY}_{th}$&$T_{1/2}[Q^{KUTY}_{th}]$ s\\ 
\hline
$^{	285	}	111	$&$	8.265	$&$	6.10E+04	$&$^{	252	}	112	$&$	14.935	$&$	1.02E-10	$\\
$^{	286	}	111	$&$	8.195	$&$	4.87E+05	$&$^{	253	}	112	$&$	18.775	$&$	6.02E-15	$\\
$^{	287	}	111	$&$	8.115	$&$	2.06E+05	$&$^{	254	}	112	$&$	13.535	$&$	1.51E-08	$\\
$^{	288	}	111	$&$	8.085	$&$	1.19E+06	$&$^{	255	}	112	$&$	13.465	$&$	4.25E-08	$\\
$^{	289	}	111	$&$	7.955	$&$	8.31E+05	$&$^{	256	}	112	$&$	13.415	$&$	2.26E-08	$\\
$^{	290	}	111	$&$	7.755	$&$	2.27E+07	$&$^{	257	}	112	$&$	13.345	$&$	6.41E-08	$\\
\hline
\hline
\end{tabular} 
\end{center}
\end{table}

\begin{table}[tbh]
%\caption{Alpha-decay half lives of nuclei using $Q$-values (MeV) from KUTY~\cite{KUTY}}
\begin{center}
\begin{tabular}{cccccc}
\hline \hline
 Parent &KUTY&DDM3Y(KUTY)&Parent &KUTY&DDM3Y(KUTY)    \\ 
\hline
$^A Z$&$Q^{KUTY}_{th}$&$T_{1/2}[Q^{KUTY}_{th}]$ s&$^A Z$&$Q^{KUTY}_{th}$&$T_{1/2}[Q^{KUTY}_{th}]$ s\\ 
\hline
$^{	279	}	112	$&$	11.025	$&$	1.78E-03	$&$^{	306	}	112	$&$	7.325	$&$	2.98E+08	$\\
$^{	280	}	112	$&$	10.995	$&$	8.41E-04	$&$^{	307	}	112	$&$	7.205	$&$	3.08E+09	$\\
$^{	281	}	112	$&$	10.685	$&$	1.11E-02	$&$^{	308	}	112	$&$	7.095	$&$	3.27E+09	$\\
$^{	282	}	112	$&$	10.135	$&$	1.18E-01	$&$^{	309	}	112	$&$	6.985	$&$	3.12E+10	$\\
$^{	283	}	112	$&$	9.595	$&$	9.84E+00	$&$^{	310	}	112	$&$	6.865	$&$	3.87E+10	$\\
$^{	284	}	112	$&$	9.225	$&$	4.82E+01	$&$^{	311	}	112	$&$	6.715	$&$	6.41E+11	$\\
\hline
\hline
\end{tabular} 
\end{center}
\end{table}

\begin{table}[tbh]
%\caption{Alpha-decay half lives of nuclei using $Q$-values (MeV) from KUTY~\cite{KUTY}}
\begin{center}
\begin{tabular}{cccccc}
\hline \hline
 Parent &KUTY&DDM3Y(KUTY)&Parent &KUTY&DDM3Y(KUTY)    \\ 
\hline
$^A Z$&$Q^{KUTY}_{th}$&$T_{1/2}[Q^{KUTY}_{th}]$ s&$^A Z$&$Q^{KUTY}_{th}$&$T_{1/2}[Q^{KUTY}_{th}]$ s\\ 
\hline
$^{	275	}	113	$&$	12.095	$&$	1.26E-05	$&$^{	302	}	113	$&$	8.855	$&$	8.37E+03	$\\
$^{	276	}	113	$&$	11.985	$&$	7.17E-05	$&$^{	303	}	113	$&$	8.555	$&$	2.07E+04	$\\
$^{	277	}	113	$&$	11.865	$&$	3.58E-05	$&$^{	304	}	113	$&$	8.305	$&$	7.40E+05	$\\
$^{	278	}	113	$&$	11.625	$&$	4.18E-04	$&$^{	305	}	113	$&$	7.965	$&$	3.26E+06	$\\
$^{	279	}	113	$&$	11.535	$&$	1.81E-04	$&$^{	306	}	113	$&$	7.485	$&$	1.53E+09	$\\
$^{	280	}	113	$&$	11.455	$&$	9.44E-04	$&$^{	307	}	113	$&$	7.455	$&$	4.23E+08	$\\
\hline
\hline
\end{tabular} 
\end{center}
\end{table}

\begin{table}[tbh]
%\caption{Alpha-decay half lives of nuclei using $Q$-values (MeV) from KUTY~\cite{KUTY}}
\begin{center}
\begin{tabular}{cccccc}
\hline \hline
 Parent &KUTY&DDM3Y(KUTY)&Parent &KUTY&DDM3Y(KUTY)    \\ 
\hline
$^A Z$&$Q^{KUTY}_{th}$&$T_{1/2}[Q^{KUTY}_{th}]$ s&$^A Z$&$Q^{KUTY}_{th}$&$T_{1/2}[Q^{KUTY}_{th}]$ s\\ 
\hline
$^{	274	}	114	$&$	12.715	$&$	9.13E-07	$&$^{	301	}	114	$&$	9.745	$&$	1.13E+01	$\\
$^{	275	}	114	$&$	12.595	$&$	3.60E-06	$&$^{	302	}	114	$&$	9.645	$&$	8.36E+00	$\\
$^{	276	}	114	$&$	12.485	$&$	2.47E-06	$&$^{	303	}	114	$&$	9.335	$&$	1.93E+02	$\\
$^{	277	}	114	$&$	12.425	$&$	7.32E-06	$&$^{	304	}	114	$&$	9.065	$&$	5.08E+02	$\\
$^{	278	}	114	$&$	12.305	$&$	5.28E-06	$&$^{	305	}	114	$&$	8.825	$&$	8.35E+03	$\\
$^{	279	}	114	$&$	12.205	$&$	1.91E-05	$&$^{	306	}	114	$&$	8.415	$&$	8.64E+04	$\\
\hline
\hline
\end{tabular} 
\end{center}
\end{table}

\begin{table}[tbh]
%\caption{Alpha-decay half lives of nuclei using $Q$-values (MeV) from KUTY~\cite{KUTY}}
\begin{center}
\begin{tabular}{cccccc}
\hline \hline
 Parent &KUTY&DDM3Y(KUTY)&Parent &KUTY&DDM3Y(KUTY)    \\ 
\hline
$^A Z$&$Q^{KUTY}_{th}$&$T_{1/2}[Q^{KUTY}_{th}]$ s&$^A Z$&$Q^{KUTY}_{th}$&$T_{1/2}[Q^{KUTY}_{th}]$ s\\ 
\hline
$^{	274	}	115	$&$	13.205	$&$	1.24E-06	$&$^{	301	}	115	$&$	10.655	$&$	5.51E-02	$\\
$^{	275	}	115	$&$	13.115	$&$	5.08E-07	$&$^{	302	}	115	$&$	10.495	$&$	5.55E-01	$\\
$^{	276	}	115	$&$	13.015	$&$	2.62E-06	$&$^{	303	}	115	$&$	10.235	$&$	6.91E-01	$\\
$^{	277	}	115	$&$	12.925	$&$	1.11E-06	$&$^{	304	}	115	$&$	9.895	$&$	2.49E+01	$\\
$^{	278	}	115	$&$	12.845	$&$	5.20E-06	$&$^{	305	}	115	$&$	9.585	$&$	4.93E+01	$\\
$^{	279	}	115	$&$	12.725	$&$	2.52E-06	$&$^{	306	}	115	$&$	9.315	$&$	1.36E+03	$\\
\hline
\hline
\end{tabular} 
\end{center}
\end{table}

\begin{table}[tbh]
%\caption{Alpha-decay half lives of nuclei using $Q$-values (MeV) from KUTY~\cite{KUTY}}
\begin{center}
\begin{tabular}{cccccc}
\hline \hline
 Parent &KUTY&DDM3Y(KUTY)&Parent &KUTY&DDM3Y(KUTY)    \\ 
\hline
$^A Z$&$Q^{KUTY}_{th}$&$T_{1/2}[Q^{KUTY}_{th}]$ s&$^A Z$&$Q^{KUTY}_{th}$&$T_{1/2}[Q^{KUTY}_{th}]$ s\\ 
\hline
$^{	276	}	116	$&$	13.485	$&$	1.27E-07	$&$^{	303	}	116	$&$	11.215	$&$	6.74E-03	$\\
$^{	277	}	116	$&$	13.415	$&$	3.84E-07	$&$^{	304	}	116	$&$	10.885	$&$	1.67E-02	$\\
$^{	278	}	116	$&$	13.355	$&$	2.02E-07	$&$^{	305	}	116	$&$	10.485	$&$	4.53E-01	$\\
$^{	279	}	116	$&$	13.285	$&$	6.22E-07	$&$^{	306	}	116	$&$	10.125	$&$	1.63E+00	$\\
$^{	280	}	116	$&$	12.965	$&$	1.02E-06	$&$^{	307	}	116	$&$	9.855	$&$	2.52E+01	$\\
$^{	281	}	116	$&$	12.705	$&$	7.49E-06	$&$^{	308	}	116	$&$	9.365	$&$	2.83E+02	$\\
\hline
\hline
\end{tabular} 
\end{center}
\end{table}

\begin{table}[tbh]
%\caption{Alpha-decay half lives of nuclei using $Q$-values (MeV) from KUTY~\cite{KUTY}}
\begin{center}
\begin{tabular}{cccccc}
\hline \hline
 Parent &KUTY&DDM3Y(KUTY)&Parent &KUTY&DDM3Y(KUTY)    \\ 
\hline
$^A Z$&$Q^{KUTY}_{th}$&$T_{1/2}[Q^{KUTY}_{th}]$ s&$^A Z$&$Q^{KUTY}_{th}$&$T_{1/2}[Q^{KUTY}_{th}]$ s\\ 
\hline
$^{	280	}	117	$&$	13.675	$&$	5.55E-07	$&$^{	307	}	117	$&$	10.685	$&$	1.93E-01	$\\
$^{	281	}	117	$&$	12.965	$&$	3.30E-06	$&$^{	308	}	117	$&$	10.405	$&$	4.26E+00	$\\
$^{	282	}	117	$&$	12.755	$&$	2.85E-05	$&$^{	309	}	117	$&$	9.825	$&$	4.53E+01	$\\
$^{	283	}	117	$&$	12.505	$&$	2.49E-05	$&$^{	310	}	117	$&$	9.185	$&$	1.95E+04	$\\
$^{	284	}	117	$&$	12.195	$&$	3.88E-04	$&$^{	311	}	117	$&$	8.865	$&$	5.54E+04	$\\
$^{	285	}	117	$&$	11.875	$&$	5.16E-04	$&$^{	312	}	117	$&$	8.715	$&$	8.13E+05	$\\
\hline
\hline
\end{tabular} 
\end{center}
\end{table}

\include{table2}
\begin{table}[tbh]
%\caption{Alpha-decay half lives of nuclei using $Q$-values (MeV) from KUTY~\cite{KUTY}}
\begin{center}
\begin{tabular}{cccccc}
\hline \hline
 Parent &KUTY&DDM3Y(KUTY)&Parent &KUTY&DDM3Y(KUTY)    \\ 
\hline
$^A Z$&$Q^{KUTY}_{th}$&$T_{1/2}[Q^{KUTY}_{th}]$ s&$^A Z$&$Q^{KUTY}_{th}$&$T_{1/2}[Q^{KUTY}_{th}]$ s\\ 
\hline
$^{	287	}	118	$&$	12.155	$&$	3.51E-04	$&$^{	314	}	118	$&$	9.035	$&$	1.76E+04	$\\
$^{	288	}	118	$&$	11.905	$&$	4.90E-04	$&$^{	315	}	118	$&$	8.925	$&$	1.13E+05	$\\
$^{	289	}	118	$&$	11.785	$&$	2.22E-03	$&$^{	316	}	118	$&$	8.815	$&$	9.56E+04	$\\
$^{	290	}	118	$&$	11.645	$&$	1.82E-03	$&$^{	317	}	118	$&$	8.705	$&$	6.46E+05	$\\
$^{	291	}	118	$&$	11.595	$&$	5.68E-03	$&$^{	318	}	118	$&$	8.575	$&$	6.60E+05	$\\
$^{	292	}	118	$&$	11.465	$&$	4.48E-03	$&$^{	274	}	119	$&$	14.975	$&$	1.61E-08	$\\
\hline
\hline
\end{tabular} 
\end{center}
\end{table}

\begin{table}[tbh]
%\caption{Alpha-decay half lives of nuclei using $Q$-values (MeV) from KUTY~\cite{KUTY}}
\begin{center}
\begin{tabular}{cccccc}
\hline \hline
 Parent &KUTY&DDM3Y(KUTY)&Parent &KUTY&DDM3Y(KUTY)    \\ 
\hline
$^A Z$&$Q^{KUTY}_{th}$&$T_{1/2}[Q^{KUTY}_{th}]$ s&$^A Z$&$Q^{KUTY}_{th}$&$T_{1/2}[Q^{KUTY}_{th}]$ s\\ 
\hline
$^{	296	}	119	$&$	11.505	$&$	4.37E-02	$&$^{	280	}	120	$&$	15.005	$&$	4.28E-09	$\\
$^{	297	}	119	$&$	11.285	$&$	3.75E-02	$&$^{	281	}	120	$&$	14.765	$&$	2.20E-08	$\\
$^{	298	}	119	$&$	11.325	$&$	1.16E-01	$&$^{	282	}	120	$&$	14.255	$&$	6.65E-08	$\\
$^{	299	}	119	$&$	11.475	$&$	1.24E-02	$&$^{	283	}	120	$&$	13.545	$&$	2.67E-06	$\\
$^{	300	}	119	$&$	11.395	$&$	7.41E-02	$&$^{	284	}	120	$&$	12.675	$&$	5.39E-05	$\\
$^{	301	}	119	$&$	11.345	$&$	2.40E-02	$&$^{	285	}	120	$&$	12.955	$&$	3.43E-05	$\\
\hline
\hline
\end{tabular} 
\end{center}
\end{table}

\begin{table}[tbh]
%\caption{Alpha-decay half lives of nuclei using $Q$-values (MeV) from KUTY~\cite{KUTY}}
\begin{center}
\begin{tabular}{cccccc}
\hline \hline
 Parent &KUTY&DDM3Y(KUTY)&Parent &KUTY&DDM3Y(KUTY)    \\ 
\hline
$^A Z$&$Q^{KUTY}_{th}$&$T_{1/2}[Q^{KUTY}_{th}]$ s&$^A Z$&$Q^{KUTY}_{th}$&$T_{1/2}[Q^{KUTY}_{th}]$ s\\ 
\hline
$^{	307	}	120	$&$	13.005	$&$	1.42E-05	$&$^{	294	}	121	$&$	13.075	$&$	7.59E-05	$\\
$^{	308	}	120	$&$	12.775	$&$	1.64E-05	$&$^{	295	}	121	$&$	12.905	$&$	4.49E-05	$\\
$^{	309	}	120	$&$	12.615	$&$	8.39E-05	$&$^{	296	}	121	$&$	12.805	$&$	2.59E-04	$\\
$^{	310	}	120	$&$	12.455	$&$	7.23E-05	$&$^{	297	}	121	$&$	12.685	$&$	1.18E-04	$\\
$^{	311	}	120	$&$	12.195	$&$	6.07E-04	$&$^{	298	}	121	$&$	12.445	$&$	1.37E-03	$\\
$^{	312	}	120	$&$	11.805	$&$	1.82E-03	$&$^{	299	}	121	$&$	12.185	$&$	1.31E-03	$\\
\hline
\hline
\end{tabular} 
\end{center}
\end{table}

\begin{table}[tbh]
%\caption{Alpha-decay half lives of nuclei using $Q$-values (MeV) from KUTY~\cite{KUTY}}
\begin{center}
\begin{tabular}{cccccc}
\hline \hline
 Parent &KUTY&DDM3Y(KUTY)&Parent &KUTY&DDM3Y(KUTY)    \\ 
\hline
$^A Z$&$Q^{KUTY}_{th}$&$T_{1/2}[Q^{KUTY}_{th}]$ s&$^A Z$&$Q^{KUTY}_{th}$&$T_{1/2}[Q^{KUTY}_{th}]$ s\\ 
\hline
$^{	321	}	121	$&$	9.875	$&$	7.07E+02	$&$^{	310	}	122	$&$	13.825	$&$	6.16E-07	$\\
$^{	284	}	122	$&$	14.315	$&$	1.79E-07	$&$^{	311	}	122	$&$	13.635	$&$	3.08E-06	$\\
$^{	285	}	122	$&$	13.795	$&$	3.39E-06	$&$^{	312	}	122	$&$	13.495	$&$	2.31E-06	$\\
$^{	286	}	122	$&$	13.505	$&$	4.78E-06	$&$^{	313	}	122	$&$	13.065	$&$	3.67E-05	$\\
$^{	287	}	122	$&$	13.945	$&$	1.72E-06	$&$^{	314	}	122	$&$	12.615	$&$	1.20E-04	$\\
$^{	288	}	122	$&$	14.455	$&$	8.93E-08	$&$^{	315	}	122	$&$	12.285	$&$	1.50E-03	$\\
\hline
\hline
\end{tabular} 
\end{center}
\end{table}

\begin{table}[tbh]
%\caption{Alpha-decay half lives of nuclei using $Q$-values (MeV) from KUTY~\cite{KUTY}}
\begin{center}
\begin{tabular}{cccccc}
\hline \hline
 Parent &KUTY&DDM3Y(KUTY)&Parent &KUTY&DDM3Y(KUTY)    \\ 
\hline
$^A Z$&$Q^{KUTY}_{th}$&$T_{1/2}[Q^{KUTY}_{th}]$ s&$^A Z$&$Q^{KUTY}_{th}$&$T_{1/2}[Q^{KUTY}_{th}]$ s\\ 
\hline
$^{	301	}	123	$&$	13.005	$&$	9.65E-05	$&$^{	295	}	124	$&$	14.675	$&$	2.58E-07	$\\
$^{	302	}	123	$&$	13.195	$&$	1.43E-04	$&$^{	296	}	124	$&$	14.565	$&$	1.66E-07	$\\
$^{	303	}	123	$&$	13.415	$&$	1.40E-05	$&$^{	297	}	124	$&$	14.485	$&$	5.07E-07	$\\
$^{	304	}	123	$&$	13.325	$&$	7.50E-05	$&$^{	298	}	124	$&$	14.335	$&$	3.85E-07	$\\
$^{	305	}	123	$&$	13.255	$&$	2.68E-05	$&$^{	299	}	124	$&$	14.125	$&$	2.11E-06	$\\
$^{	306	}	123	$&$	13.195	$&$	1.27E-04	$&$^{	300	}	124	$&$	13.955	$&$	1.75E-06	$\\
\hline
\hline
\end{tabular} 
\end{center}
\end{table}

\begin{table}[tbh]
%\caption{Alpha-decay half lives of nuclei using $Q$-values (MeV) from KUTY~\cite{KUTY}}
\begin{center}
\begin{tabular}{cccccc}
\hline \hline
 Parent &KUTY&DDM3Y(KUTY)&Parent &KUTY&DDM3Y(KUTY)    \\ 
\hline
$^A Z$&$Q^{KUTY}_{th}$&$T_{1/2}[Q^{KUTY}_{th}]$ s&$^A Z$&$Q^{KUTY}_{th}$&$T_{1/2}[Q^{KUTY}_{th}]$ s\\ 
\hline
$^{	322	}	124	$&$	12.165	$&$	3.94E-03	$&$^{	319	}	125	$&$	13.285	$&$	6.34E-05	$\\
$^{	323	}	124	$&$	11.975	$&$	2.62E-02	$&$^{	320	}	125	$&$	12.945	$&$	1.14E-03	$\\
$^{	324	}	124	$&$	14.235	$&$	2.91E-07	$&$^{	321	}	125	$&$	12.765	$&$	7.10E-04	$\\
$^{	295	}	125	$&$	15.385	$&$	2.46E-08	$&$^{	322	}	125	$&$	12.655	$&$	4.43E-03	$\\
$^{	296	}	125	$&$	15.255	$&$	1.28E-07	$&$^{	323	}	125	$&$	12.515	$&$	2.29E-03	$\\
$^{	297	}	125	$&$	15.125	$&$	5.97E-08	$&$^{	324	}	125	$&$	12.385	$&$	1.64E-02	$\\
\hline
\hline
\end{tabular} 
\end{center}
\end{table}

\begin{table}[tbh]
%\caption{Alpha-decay half lives of nuclei using $Q$-values (MeV) from KUTY~\cite{KUTY}}
\begin{center}
\begin{tabular}{cccccc}
\hline \hline
 Parent &KUTY&DDM3Y(KUTY)&Parent &KUTY&DDM3Y(KUTY)    \\ 
\hline
$^A Z$&$Q^{KUTY}_{th}$&$T_{1/2}[Q^{KUTY}_{th}]$ s&$^A Z$&$Q^{KUTY}_{th}$&$T_{1/2}[Q^{KUTY}_{th}]$ s\\ 
\hline
$^{	318	}	126	$&$	14.195	$&$	1.43E-06	$&$^{	320	}	127	$&$	14.205	$&$	1.51E-05	$\\
$^{	319	}	126	$&$	13.845	$&$	1.49E-05	$&$^{	321	}	127	$&$	14.105	$&$	6.27E-06	$\\
$^{	320	}	126	$&$	13.765	$&$	8.17E-06	$&$^{	322	}	127	$&$	13.675	$&$	1.46E-04	$\\
$^{	321	}	126	$&$	13.315	$&$	1.47E-04	$&$^{	323	}	127	$&$	13.545	$&$	6.66E-05	$\\
$^{	322	}	126	$&$	13.165	$&$	1.16E-04	$&$^{	324	}	127	$&$	13.415	$&$	4.49E-04	$\\
$^{	323	}	126	$&$	13.015	$&$	5.81E-04	$&$^{	325	}	127	$&$	13.245	$&$	2.59E-04	$\\
\hline
\hline
\end{tabular} 
\end{center}
\end{table}

\begin{table}[tbh]
%\caption{Alpha-decay half lives of nuclei using $Q$-values (MeV) from KUTY~\cite{KUTY}}
\begin{center}
\begin{tabular}{cccccc}
\hline \hline
 Parent &KUTY&DDM3Y(KUTY)&Parent &KUTY&DDM3Y(KUTY)    \\ 
\hline
$^A Z$&$Q^{KUTY}_{th}$&$T_{1/2}[Q^{KUTY}_{th}]$ s&$^A Z$&$Q^{KUTY}_{th}$&$T_{1/2}[Q^{KUTY}_{th}]$ s\\ 
\hline
$^{	325	}	128	$&$	13.775	$&$	6.60E-05	$&$^{	327	}	129	$&$	13.965	$&$	3.65E-05	$\\
$^{	326	}	128	$&$	13.635	$&$	4.70E-05	$&$^{	328	}	129	$&$	13.835	$&$	2.31E-04	$\\
$^{	327	}	128	$&$	13.495	$&$	2.19E-04	$&$^{	329	}	129	$&$	13.725	$&$	9.98E-05	$\\
$^{	328	}	128	$&$	13.325	$&$	1.89E-04	$&$^{	313	}	130	$&$	17.325	$&$	5.73E-10	$\\
$^{	309	}	129	$&$	17.165	$&$	4.25E-10	$&$^{	314	}	130	$&$	17.185	$&$	3.76E-10	$\\
$^{	310	}	129	$&$	17.085	$&$	1.75E-09	$&$^{	315	}	130	$&$	17.965	$&$	7.93E-11	$\\
\hline
\hline
\end{tabular} 
\end{center}
\end{table}

\end{document}